\documentclass[aps,prl,twocolumn,showpacs]{revtex4}

\usepackage{epsfig}
\usepackage{graphicx}
\usepackage{amsmath}
\usepackage{bm}

\begin{document}

\title{Magnetic correlations in the Hubbard model
on triangular and Kagom\'{e} lattices}

\author{N. Bulut$^{1,2}$, W. Koshibae$^1$ and S. Maekawa$^{1,2}$}

\address{ $^1$Institute for Materials Research, Tohoku University, Sendai
980-8577, Japan \\
$^2$CREST, Japan Science and Technology Agency (JST), Kawaguchi,
Saitama 332-0012, Japan }

\date{February 15, 2005}

\begin{abstract}
In order to study the magnetic properties of frustrated metallic
systems, we present Quantum Monte Carlo data on the magnetic
susceptibility of the Hubbard model on triangular and Kagom\'{e}
lattices. We show that the underlying lattice structure is
important, and determines the nature and the doping dependence of
the magnetic fluctuations. In particular, in the doped Kagom\'{e}
case we find strong short-range magnetic correlations, which makes
the metallic Kagom\'{e} systems a promising field for studies of
superconductivity.
\end{abstract}


\maketitle

Frustrated spin systems have received significant attention
because of the possibility of novel magnetic ground-states and
excitations \cite{Misguich}. The triangular spin-$1/2$ Heisenberg
model has long-range magnetic order in the ground state, while the
spin-$1/2$ Heisenberg model on the Kagom\'{e} lattice is
considered to be disordered. The discovery of superconductivity at
$5 K$ in Na$_x$CoO$_2\cdot y$H$_2$O has generated new interest in
frustrated interacting systems\cite{Takada}. Furthermore, recently
superconductivity has been discovered in $\beta$-pyrochlore osmate
KOs$_2$O$_6$ with $T_c$ of $10K$ \cite{Yonezawa}. This is
interesting since the pyrochlore lattice structure is a
three-dimensional analog of the Kagome lattice.

In cobaltates, cobalt and oxygen ions form a two-dimensional
triangular network. The hopping matrix element of electrons in the
cobalt $3d$ orbitals is not isotropic, and it has been shown that
the triangular CoO$_2$ lattice consists of four coupled Kagom\'{e}
sublattices \cite{Koshibae}. Hence, it is important to compare the
magnetic properties of interacting systems on triangular and
Kagom\'{e} lattices. The electronic properties of the $t$-$J$ and
the Hubbard models on the triangular and Kagom\'{e} lattices have
been studied using various techniques of many-body physics. The
triangular $t$-$J$ model was investigated within the RVB framework
\cite{Baskaran,Kumar,Wang} and by using high-temperature
expansions \cite{Koretsune}. The triangular Hubbard model was
studied with the path-integral renormalization-group (RG)
\cite{Kashima}, the one-loop RG \cite{Honerkamp} and the
fluctuation-exchange (FLEX) \cite{Renner} approaches. The FLEX
method was also used for studying the magnetic properties of the
Hubbard model on the Kagom\'{e} lattice \cite{Imai}.

In this paper, we compare the nature of the magnetic correlations
in the Hubbard model on the triangular and the Kagom\'{e} lattices
using Quantum Monte Carlo (QMC) simulations. We consider the
Hubbard model on the Kagom\'{e} lattice to be a simple limiting
case to explore for new physics due to the underlying orbital
structure in frustrated interacting systems. The orbital degrees
of freedom create the possibility for mixing the spin and charge
channels, and hence of new electronic states. This is important
because of the general ongoing research effort on the transition
metal oxides. These are our motivations for performing the QMC
calculations on the Hubbard model on the triangular and the
Kagom\'{e} lattices. We are particularly interested in the doped
cases of these models for which there are no exact calculations on
the magnetic properties.

In the following, we will see that the nature of the magnetic
fluctuations on the triangular and the Kagom\'{e} lattices are
different. For the triangular lattice, the QMC results show that
there are strong antiferromagnetic (AF) correlations near
half-filling at low temperatures, when the Coulomb repulsion $U$
is of the order of the bandwidth. On the other hand, for weak $U$,
the magnetic correlations saturate as $T\rightarrow 0$. In the
Kagom\'{e} lattice, the unit cell consists of three atoms and the
unit cells form a triangular lattice, as seen in Fig. 1(a).
Consequently, there are three bands of magnetic excitations. Two
of these modes involve enhanced short-range AF correlations. We
find that, in the doped Kagom\'{e} case, the low-frequency
short-range AF correlations are stronger in comparison to the
triangular lattice. Hence, we note that it would be useful to
investigate the possibility of superconductivity in metallic
Kagom\'{e} systems.
\begin{figure}[b]
\begin{center}
\epsfig{file=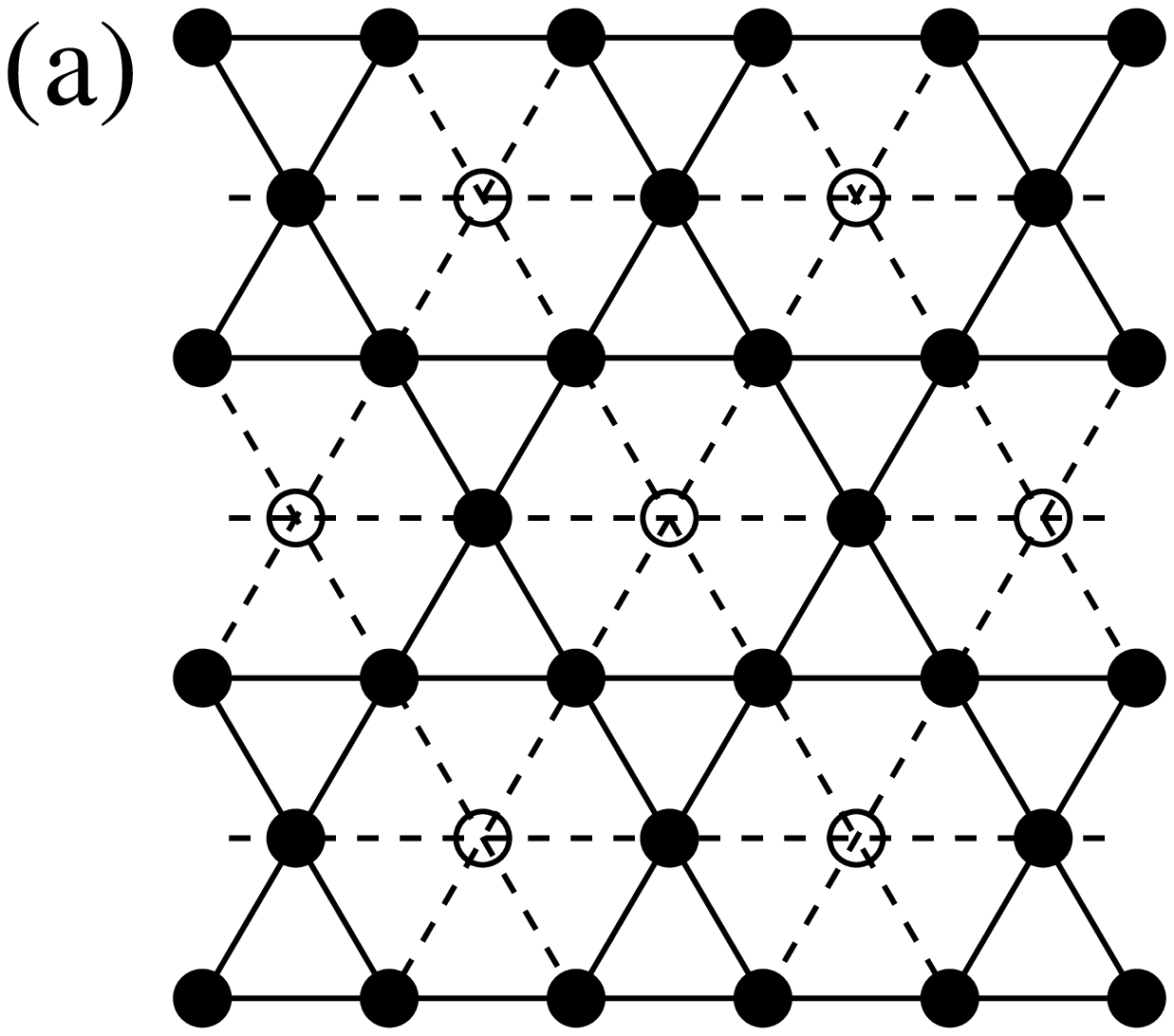,height=3cm}
\epsfig{file=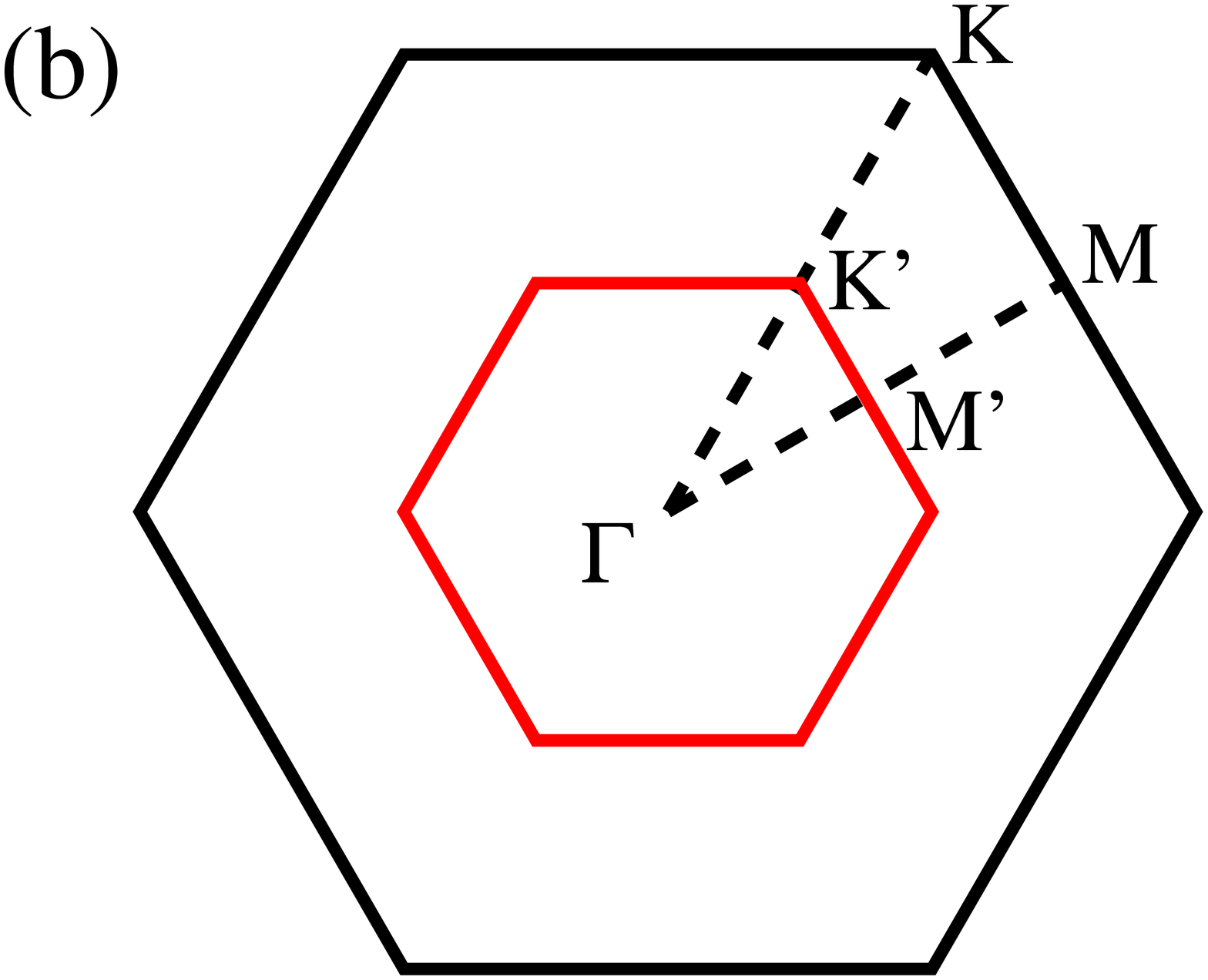,height=3cm}
\end{center}
\caption{ (a) Sketch of the triangular and the Kagom\'{e}
lattices. The Kagom\'{e} lattice is obtained from the triangular
lattice by removing the sites with the open circles. (b) First
Brillouin zones of the triangular (black) and the Kagom\'{e} (red)
lattices. } \label{fig1}
\end{figure}

The Hubbard model is defined by
\begin{eqnarray}
\label{Hubbard} H=-t \sum_{\langle i,j\rangle,\sigma}
(c^{\dagger}_{i\sigma}c_{j\sigma} + {\rm h.c.}) + U \sum_{i}
n_{i\uparrow} n_{i\downarrow} -\mu\sum_{i\sigma}n_{i\sigma},
\end{eqnarray}
where $t$ is the hopping matrix element between the nearest
neighbor-sites, $U$ is the on-site Coulomb repulsion, and $\mu$ is
the chemical potential. Here, $c_{i\sigma}$
($c^{\dagger}_{i\sigma}$) annihilates (creates) an electron with
spin $\sigma$ at site $i$, and $n_{i\sigma}=c^{\dagger}_{i\sigma}
c_{i\sigma}$. In the following, we will take $t<0$ and consider
$\langle n\rangle \ge 1.0$, which is the appropriate case for the
cobaltates \cite{Koshibae}. In obtaining the QMC data presented
here, the determinantal QMC technique \cite{White} was used.

For the triangular lattice, the longitudinal magnetic
susceptibility at frequency $\omega=0$ is defined by
\begin{equation}
\chi({\bf q}) = \int_0^{\beta} d\tau \sum_{\ell} e^{-i{\bf
q}\cdot{\bf r}_{\ell}} \langle m^z({\bf r}_{i+\ell},\tau) m^z({\bf
r}_i) \rangle,
\end{equation}
where $m^z({\bf r}_i) = c^{\dagger}_{i\uparrow}c_{i\uparrow} -
c^{\dagger}_{i\downarrow}c_{i\downarrow}$ and $m^z({\bf r}_i,\tau)
= e^{H\tau} m^z({\bf r}_i) e^{-H\tau}$. In this paper, results on
$\chi$ will be shown in units of $|t|^{-1}$.

The Kagom\'{e} lattice is a three-band model, since each unit cell
consists of three sites. Hence, each lattice on the Kagom\'{e}
lattice site can be represented by the indices $(\ell,d)$ where
$\ell$ is the unit-cell index and $d$ denotes the atomic site in a
particular unit cell. The corresponding Brillouin zone (BZ) is
reduced with respect to the triangular case as illustrated in Fig.
1(b). For the Kagom\'{e} case, we define the longitudinal magnetic
susceptibility at $\omega=0$ as
\begin{equation}
\chi_{dd'}({\bf q}) = \int_0^{\beta} d\tau \sum_{\ell} e^{-i{\bf
q}\cdot{\bf r}_{\ell}} \langle m^z_d({\bf r}_{i+\ell},\tau)
m^z_{d'}({\bf r}_i) \rangle,
\end{equation}
where $m^z_d({\bf r}_i) = c^{\dagger}_{id\uparrow}c_{id\uparrow} -
c^{\dagger}_{id\downarrow}c_{id\downarrow}$, $c_{id\sigma}$
($c^{\dagger}_{id\sigma}$) is the annihilation (creation) operator
of an electron with spin $\sigma$ at lattice site $(i,d)$ and the
summation is performed over the unit-cell locations. Diagonalizing
the $3\times 3$ matrix $\chi_{dd'}({\bf q})$, we obtain
$\chi_{\alpha}({\bf q})$ which describes the three modes of the
magnetic excitations on the Kagom\'{e} lattice, which we define as
the Kagom\'{e} magnetic bands. At this point, it is useful to note
that already for the noninteracting ($U=0$) case, the triangular
and the Kagom\'{e} lattices have different properties. For $t< 0$,
the one-electron density of states $N(\omega)$ of the triangular
lattice has a van Hove singularity at $\langle n\rangle=0.5$. For
the Kagom\'{e} lattice, there is a $\delta$-function singularity
in $N(\omega)$ at the bottom of the band and there are van Hove
singularities at $\langle n\rangle=1.16$ and 1.51. In addition,
$N(\omega=0)$ vanishes at $\langle n\rangle=1.33$. These features
of $N(\omega)$ are also reflected in the magnetic susceptibilities
of the noninteracting case.

We first present results for the triangular lattice at
half-filling. Figure 2(a) shows $\chi({\bf q})$ versus ${\bf q}$
for $U=4|t|$ on various size lattices as the temperature is
lowered. Here, it is seen that $\chi({\bf q})$ has a broad peak
centered at the $K$ point of the BZ, and hence the system exhibits
short-range AF correlations. We also note that $\chi({\bf q})$
does not vary significantly with $T$, in particular for $0.25|t|
\le T \le 0.17|t|$. For comparison, $\chi_0({\bf q})$ for the
noninteracting system at $T=0.17|t|$ is shown by the dotted curve.
Figure 2(b) displays $\chi({\bf q})$ versus ${\bf q}$ for $U=8|t|$
at half-filling, where we observe a large Stoner enhancement of
the AF correlations. In contrast with the $U=4|t|$ case, here,
$\chi({\bf q})$ at the $K$ point grows rapidly with a Curie-like
$T$ dependence, as $T$ decreases from $1|t|$ to $0.33|t|$.
However, it is not known whether $\chi({\bf q})$ saturates at
lower $T$ for $U=8|t|$. These results show that the $T$ dependence
of $\chi({\bf q})$ depends strongly on the value of $U/|t|$ in the
triangular Hubbard model, in agreement with the findings of the
path-integral RG calculations \cite{Kashima}.
\begin{figure}[t]
\begin{center}
\leavevmode \epsfig{file=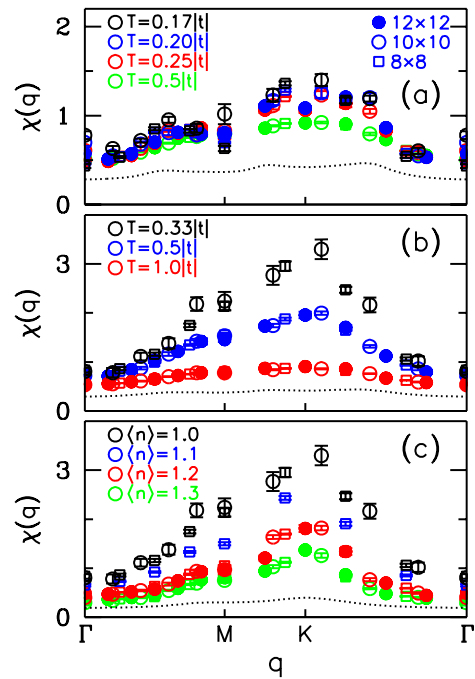, height=9cm}
\end{center}
\caption{ Magnetic susceptibility $\chi({\bf q})$ versus ${\bf q}$
of the triangular Hubbard model at zero frequency. Here, ${\bf q}$
is scanned along the path $\Gamma\rightarrow M\rightarrow
K\rightarrow \Gamma$ in the BZ of the triangular lattice
illustrated in Fig. 1(b). The temperature evolution of $\chi({\bf
q})$ at half-filling is shown in (a) for $U=4|t|$ and in (b) for
$U=8|t|$. In these figures, the dotted curves represent results
for the noninteracting case at the lowest temperature used in that
figure. The evolution of $\chi({\bf q})$ versus ${\bf q}$ with the
electron density $\langle n\rangle$ is shown in (c) for $U=8|t|$
and $T=0.33|t|$. Here, the dotted curve represents the results for
the noninteracting system at $\langle n\rangle=1.3$ and
$T=0.33|t|$. } \label{fig2}
\end{figure}

Figure 2(c) shows the filling dependence of $\chi({\bf q})$ for
$U=8|t|$ while $T$ is kept fixed at $0.33|t|$. Here, we observe
that the AF correlations decay monotonically as the electron
filling is varied from 1.0 to 1.3. We have also performed
calculations for $\chi({\bf q})$ at higher electron fillings. We
find that, when $\langle n\rangle$ is increased to 1.5, the peak
in $\chi({\bf q})$ shifts to the $M$ point. Upon further doping to
$\langle n\rangle=1.75$, we find that, for $U=8|t|$ and
$T=0.2|t|$, the Stoner enhancement is about 20\%. Hence, for this
dilute hole concentration, the magnetic correlations are weakly
affected by the presence of the on-site Coulomb repulsion at this
temperature. For $U=4|t|$, $\chi({\bf q})$ exhibits a slow
monotonic decrease with the electron doping away from
half-filling.
\begin{figure}[t]
\begin{center}
\leavevmode \epsfig{file=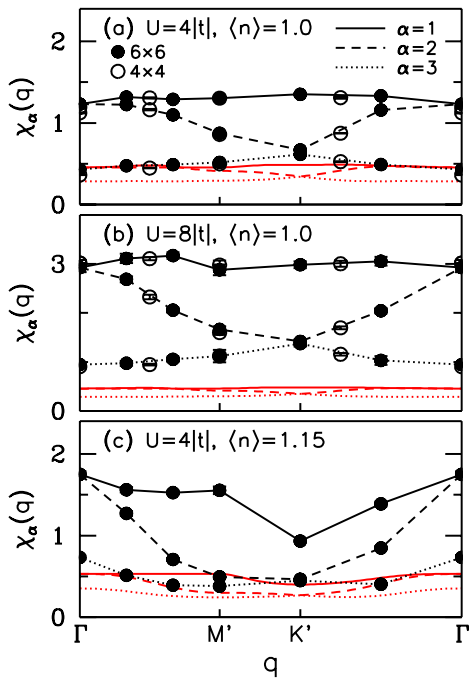, height=9cm}
\end{center}
\caption{ Magnetic susceptibility $\chi_{\alpha}({\bf q})$ versus
${\bf q}$ for the Hubbard model on the Kagom\'{e} lattice at zero
frequency. Here, the three magnetic modes of $\chi_{\alpha}({\bf
q})$ are shown at each ${\bf q}$ point, as ${\bf q}$ is scanned
along the path $\Gamma\rightarrow M'\rightarrow K'\rightarrow
\Gamma$ in the first BZ of the Kagom\'{e} lattice illustrated in
Fig. 1(b). Results on $\chi_{\alpha}({\bf q})$ at half-filling are
shown in (a) for $U=4|t|$ and $T=0.2|t|$ and in (b) for $U=8|t|$
and $T=0.33|t|$. These data have been obtained on lattices with
$6\times 6$ and $4\times 4$ unit cells. In (c), results are shown
for $\langle n\rangle =1.15$ with $U=4|t|$ and $T=0.14|t|$. In
these figures, the data points for the $6\times 6$ lattice are
connected by black lines, and the red curves represent the results
for the noninteracting case. } \label{fig3}
\end{figure}

Next, we discuss the magnetic properties of the Hubbard model on
the Kagom\'{e} lattice. Figure 3(a) shows QMC results on
$\chi_{\alpha}({\bf q})$ for $U=4|t|$ and $T=0.2|t|$ at
half-filling. Here, $\chi_{\alpha}({\bf q})$ for the three
magnetic bands are plotted as a function of ${\bf q}$, and the red
curves represent results for the noninteracting case. In this
figure, we observe that the top band ($\alpha=1$) is flat in ${\bf
q}$ space, and the second magnetic band ($\alpha=2$) is degenerate
with the first one at the zone center. The third mode is weaker in
magnitude and exhibits a smooth ${\bf q}$ dependence. Figure 3(b)
shows $\chi_{\alpha}({\bf q})$ versus ${\bf q}$ for $U=8|t|$ and
$T=0.33|t|$ at half-filling. The general features are similar to
those seen in Fig. 3(a), however here the Stoner enhancement is
larger. In Fig. 3(c), the QMC results are shown for $U=4|t|$ and
$T=0.14|t|$ at $\langle n\rangle=1.15$. These figures show that
the features of $\chi_{\alpha}({\bf q})$ are in correspondence
with those of the noninteracting case.

In order to gain insight into the origin of the Kagom\'{e}
magnetic bands, we note that the Kagom\'{e} lattice is obtained
from the triangular lattice, as illustrated in Fig. 1(a), by
removing the sites with the empty circles. This is equivalent to
putting an infinitely repulsive one-electron potential at these
sites. Bragg scattering from this static charge-density-wave field
then folds the BZ of the triangular lattice, and also mixes the
different wavevector components of the spin fluctuations. It is
this process which creates the Kagom\'{e} magnetic bands.
\begin{figure}[t]
\begin{center}
\leavevmode \epsfxsize=4.0cm \epsfysize=5.0cm \epsffile[130 130
530 550]{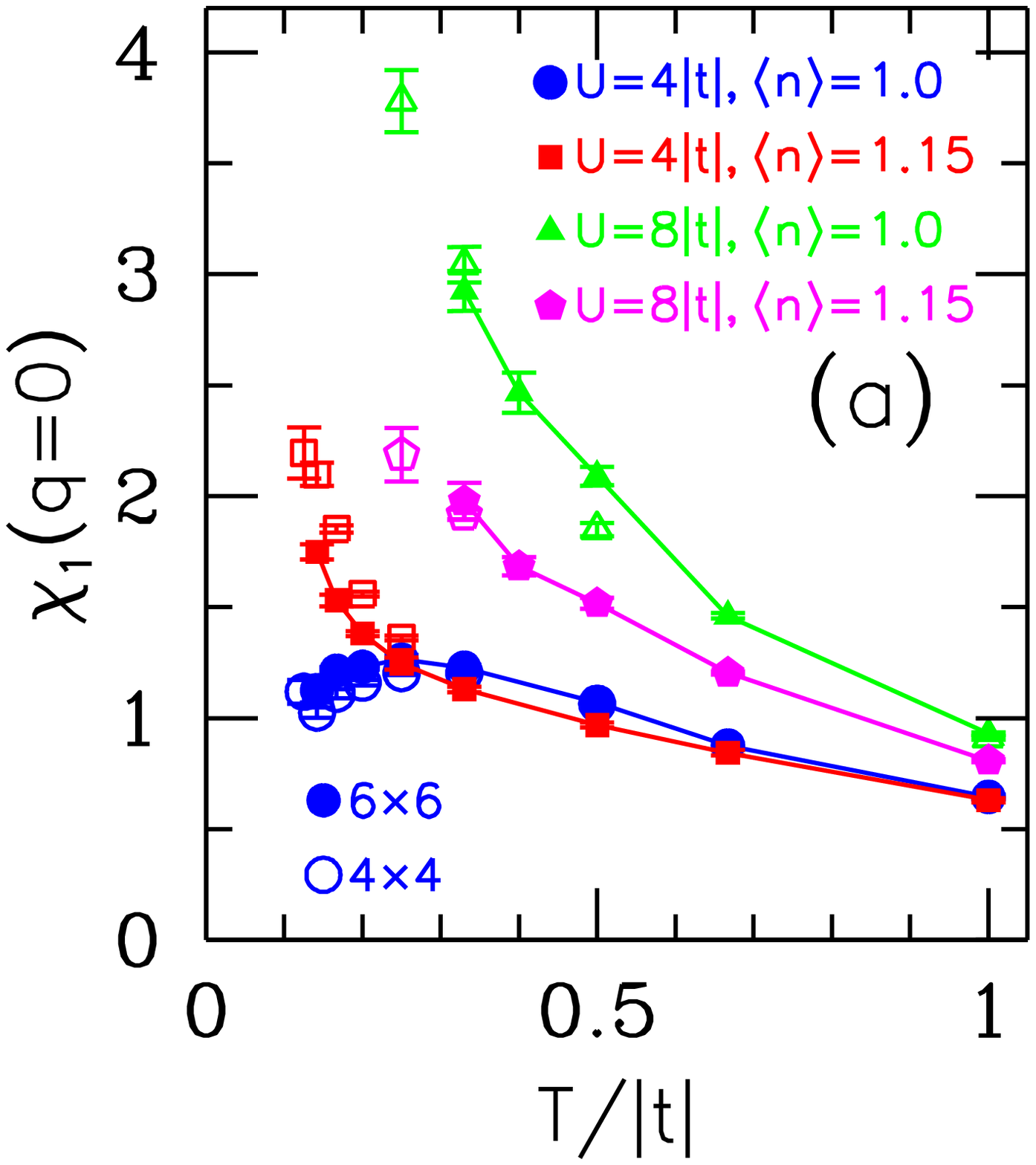} \leavevmode \epsfxsize=4.0cm \epsfysize=5.0cm
\epsffile[100 130 500 550]{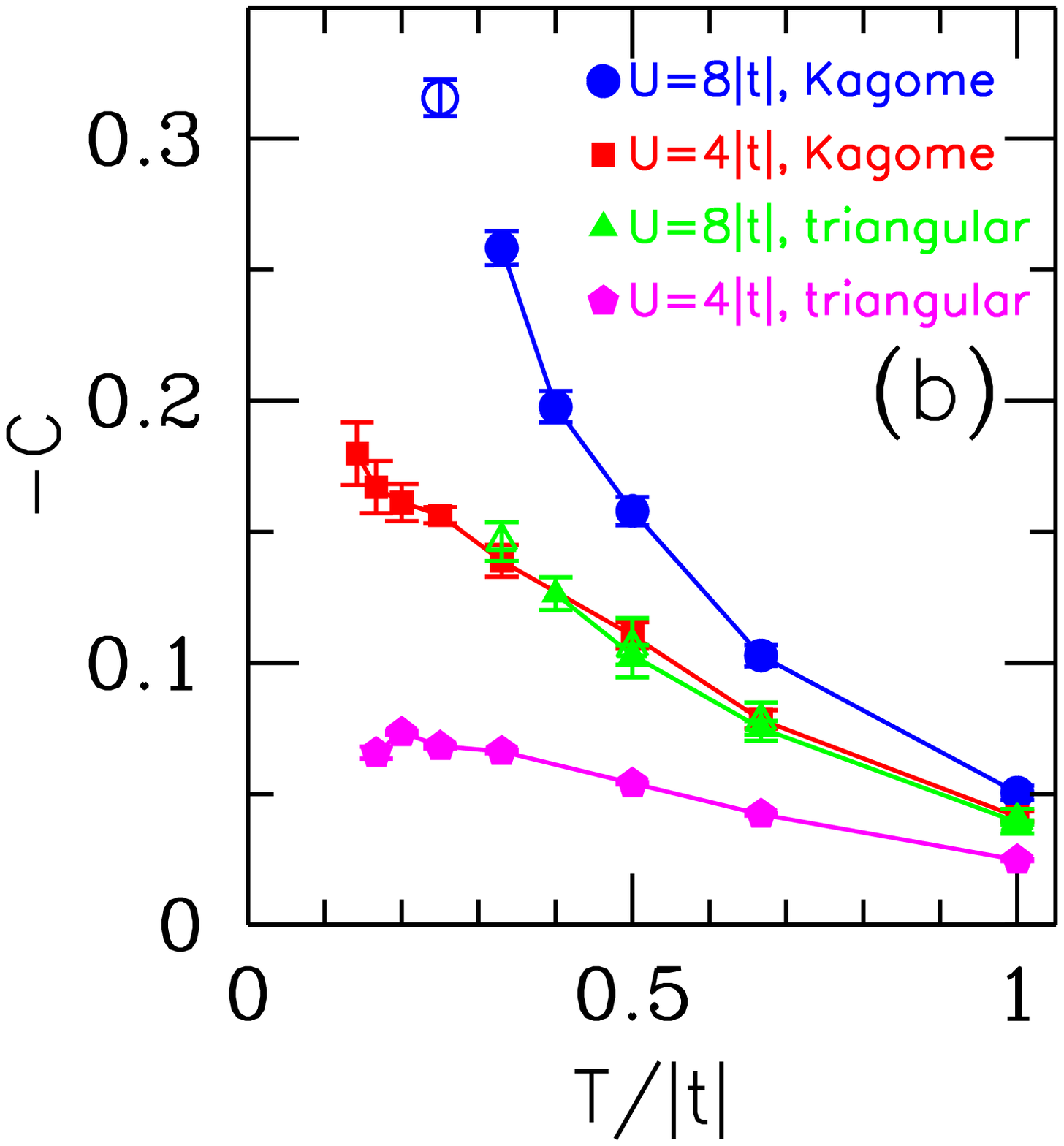}
\end{center}
\caption{ (a) $\chi_1({\bf q}=0)$ versus $T$ for the Kagom\'{e}
lattice with $U=4|t|$ and $8|t|$ at $\langle n\rangle=1.0$ and
1.15. (b) Zero-frequency component of the magnetic correlations
between the nearest-neighbor sites $C$ versus $T$ for the
Kagom\'{e} and the triangular lattices at $\langle n\rangle=1.15$
for $U=8|t|$ and $4|t|$. In these figures, the results on the
Kagom\'{e} lattice were obtained for $6\times 6$ (solid points)
and $4\times 4$ (open points) unit cells, and the results on the
triangular lattice are for $12\times 12$ (solid points) and
$8\times 8$ (empty points) lattices. In addition, the data points
for the $6\times 6$ Kagom\'{e} and the $12\times 12$ triangular
lattices have been connected by solid lines. } \label{fig4}
\end{figure}

We next discuss the $T$ dependence of the magnetic correlations in
the Kagom\'{e} lattice. Figure 4(a) shows $\chi_1({\bf q}=0)$
versus $T$ for $U=4|t|$ and $8|t|$ at $\langle n\rangle=1.0$ and
1.15. Here, we observe that, at half-filling and for $U=4|t|$,
$\chi_1({\bf q}=0)$ saturates as $T$ decreases. On the other hand,
for $U=8|t|$, $\chi_1({\bf q}=0)$ has a strong $T$-dependence. For
$\langle n\rangle=1.15$ and $U=4|t|$, we observe that $\chi_1({\bf
q}=0)$ gets enhanced at low $T$, and becomes larger than at
half-filling. We also see that the enhancement of $\chi_1({\bf
q}=0)$ depends on the lattice size. However, this type of
non-monotonic doping dependence was not observed for the
triangular Hubbard model. In addition, we observe that, for
$U=8|t|$ and at half-filling, $\chi_1({\bf q}= 0)$ exhibits a
Curie-like $T$ dependence for $0.25|t| \le T \le 1.0|t|$. In the
ground state of the Heisenberg model on the Kagom\'{e} lattice, it
is considered that a spin gap $\Delta_S\approx J/20$ exists, where
$J$ is the magnetic exchange \cite{Misguich}. For $U=8|t|$, we
have $J\approx 4t^2/U=0.5|t|$, which gives $\Delta_S\approx
0.025|t|$. We expect that $\chi_{\alpha}({\bf q})$ saturates
before $T$ becomes comparable to $\Delta_S$. However, already at
$T=0.33|t|$ we observe enhanced magnetic correlations.

The FLEX calculations for the Hubbard model on the Kagom\'{e}
lattice find that the leading magnetic mode is nearly ${\bf q}$
independent, and the tendency to electronic instabilities is
suppressed \cite{Imai}. We note that while the $\alpha=1$ mode is
nearly ${\bf q}$ independent, the $\alpha=2$ mode has
ferromagnetic ${\bf q}$ dependence in the sense that it decreases
away from the $\Gamma$ point. In addition, the eigenvectors of
$\chi_{dd'}({\bf q})$ show that, at ${\bf q}\approx 0$, the
$\alpha=1$ and 2 modes describe excitations involving the AF
polarization of the spins within a unit cell \cite{footnote}. Away
from ${\bf q}=0$, these modes have additional structures, however
they involve AF polarizations over most of the BZ. Hence, these
two modes contain enhanced short-range AF fluctuations. Within the
context of superconductivity mediated by magnetic fluctuations, an
important quantity is the zero-frequency component of the magnetic
fluctuations between the two nearest-neighbor sites $i$ and $j$,
\begin{equation}
C=\int_0^{\beta} \, d\tau \, \langle m^z({\bf r}_i,\tau) m^z({\bf
r}_j) \rangle.
\end{equation}
In Fig. 4(b), we compare C versus $T$ for the triangular and the
Kagom\'{e} lattices for $U=4|t|$ and $8|t|$ at $\langle
n\rangle=1.15$. This figure shows that, at these temperatures, the
nearest-neighbor AF correlations are stronger for the Kagome
lattice, even though the ground-state of the spin-$1/2$ Heisenberg
model has long-range order on the triangular lattice and it is
disordered in the Kagom\'{e} case. Hence, it would be useful to
investigate the possibility of superconductivity in metallic
Kagom\'{e} systems.

Finally, we discuss the implications of these data for the
magnetic correlations observed in the cobaltates. The cobaltates
have a rich phase diagram with superconductivity found for
$x\approx 0.35$ in Na$_x$CoO$_2\cdot y$H$_2$O \cite{Takada} and
with magnetic order and large quasi-particle renormalizations
observed in Na$_x$CoO$_2$ when $x$ is  near 0.75 \cite{Foo}. For
the triangular Hubbard model, we find that the magnetic
correlations are strongest at half-filling, and the correlation
effects are weak in the overdoped regime $\langle n\rangle \ge
1.5$. In contrast, for the Kagom\'{e} lattice we have seen that
the magnetic correlations can be stronger in the doped case when
$U=4|t|$. However, in both of these models, we find that the
correlation effects are most prominent in the vicinity of
half-filling at the temperatures where the QMC calculations were
performed. We note that it would be useful to determine
experimentally the ${\bf q}$ dependence of the magnetic
fluctuations in the superconducting Na$_x$CoO$_2\cdot y$H$_2$O.

In this paper, we have investigated the nature of the magnetic
correlations in the Hubbard model on the triangular and Kagom\'{e}
lattices. At the temperatures where the QMC calculations were
performed, we find, in both of these models, that the magnetic
correlations grow rapidly as $T$ decreases at half-filling for
$U=8|t|$, while they saturate when $U=4|t|$. In the triangular
Hubbard model, the AF correlations decay monotonically with the
electron doping. In the Kagom\'{e} case, on the other hand, we
have seen that the magnetic correlations can be stronger in the
doped case when $U=4|t|$. We have also seen that in the Kagom\'{e}
case the BZ is reduced and there are three modes of magnetic
excitations. The two leading modes involve short-range AF
correlations. In particular, we find that the low-frequency
short-range AF correlations are stronger in the doped Kagom\'{e}
case than in the triangular case. This makes the interacting
metallic systems with Kagom\'{e} type of lattice structures a
promising field for studies of superconductivity. We conclude that
in frustrated interacting systems the underlying lattice and
orbital structures are important in determining the magnetic
properties.

The authors are grateful to Y.Y. Bang, C. Honerkamp, M. Imada, K.
Ishida, S. Ishihara, T. Koretsune, P. Lee, Y. Motome, B. Normand,
T.M. Rice, T. Tohyama and G.-q. Zheng for helpful discussions. One
of us (N.B.) would like to thank the International Frontier Center
for Advanced Materials at Tohoku University for its kind
hospitality, and gratefully acknowledges partial support from the
Turkish Academy of Sciences through the GEBIP program
(EA-TUBA-GEBIP/2001-1-1). This work was supported by
Priority-Areas Grants from the Ministry of Education, Science,
Culture and Sport of Japan, NAREGI Japan and NEDO.


\begin{thebibliography}{999}
\bibitem{footnote} We also note that, at ${\bf q}\approx 0$, the $\alpha=3$ mode
corresponds to fluctuations involving the uniform polarization of
the spins within a unit cell, and hence the Knight shift for the
Kagom\'{e} lattice is determined by $\chi_3({\bf q}\rightarrow
0)$.
\end{thebibliography}
\end{document}